\documentstyle[aps,psfig]{revtex}
\newcommand{\be}{\begin{equation}}
\newcommand{\ee}{\end{equation}}
\newcommand{\half}{\frac{1}{2}}
\newcommand{\bea}{\begin{eqnarray}}
\newcommand{\eea}{\end{eqnarray}}
\newcommand{\nn}{\nonumber}

\newcommand{\k}{\vec{k}}

\newcommand{\D}{\mbox{D}}
\begin{document}
\preprint{ITFA-98-44}
\draft
\title{Effective Classical Theory for Real-Time SU($N$) Gauge  
Theories at High Temperature}
\author{B. J. Nauta\footnote{E-mail: nauta@wins.uva.nl} and Ch. G. van Weert}
\address{Institute for Theoretical Physics, University of Amsterdam\\
        Valckenierstraat 65, 1018 XE Amsterdam, The Netherlands}

\maketitle
\vspace{1cm}
\begin{abstract}
We derive an effective classical theory for real-time SU($N$) gauge theories
at high temperature. By separating off and integrating out
quantum fluctuations we obtain a 3D 
classical path integral over the initial fields and conjugate momenta. 
The leading hard mode contribution is incorporated in the equations of motion 
for the classical fields. This yields the gauge invariant hard thermal loop 
(HTL) effective equation of motion. No gauge-variant terms are generated 
as in treatments with an intermediate momentum cut-off. Quantum corrections 
to classical correlation functions can be calculated perturbatively. The 4D 
renormalizability of the theory ensures that the 4D counterterms are 
 sufficient to render the theory finite. The HTL contributions of the 
quantum fluctuations
provide the counterterms for the linear divergences in the classical theory.
\end{abstract}

\section{Introduction}
To understand the processes that have played a role in the early universe,
it is important to study the behavior of thermal field theory at 
high temperatures. Some of the processes with cosmological 
implications, like sphaleron transitions and the 
dynamics of weak first-order transitions, are sensitive 
to the soft modes (\(k\sim g^2T\), with \(g\) the gauge coupling)
of the magnetic sector in hot gauge theories \cite{arnold,moore}. 
This low-momentum behavior is non-perturbative, 
due to the IR problems of massless field theories in 3D \cite{arnold,linde}. 

For static magnetic fluctuations the dominant IR behavior can be isolated 
from the full quantum theory by the technique of dimensional reduction 
\cite{braaten,kajantie}, which allows one 
to study the non-perturbative behaviour by standard lattice simulations or
other imaginary-time methods \cite{lattice,gapeq}. 
In a
number of papers it has already been argued that real-time  thermal field 
theory in the high-$T$ limit reduces to a classical statistical theory
\cite{scalclas} 
Such 
a theory consists of equations of motion that govern the time-evolution
of the classical fields and an averaging over initial fields with a thermal 
weight. The goal of this paper is to obtain an effective classical theory by
integrating out quantum fluctuations using the method similar to that of
dimensional reduction \cite{nauta}.

A major problem for gauge theories is that the 
perturbative fluctuations (hard modes) do not cleanly decouple from the soft
modes one is interested in. Even in lowest order one will have to take the 
hard-thermal-loop (HTL) 
effects into account \cite{arnold,bodeker}. To handle this in a gauge
invariant manner we will propose in the present paper a way of splitting 
the field in perturbative and non-perturbative modes, other than the
standard method of introducing an intermediate scale to separate the soft 
from the hard modes \cite{bodeker,son,hu,iancu}.
In this way the influence action obtained by integrating out the hard modes 
is BRS invariant, in the HTL approximation this implies gauge invariance
\cite{kobes}. 

the theory that we obtain after the classical approximation is very similar 
to the effective theory proposed by Iancu \cite{iancu}. It consists of the
HTL equation of motion and a thermal average over initial conditions.
The approach that is taken here shows that the HTL's should be calculated 
with a subtracted propagator. The subtraction follows directly from separating
off the path integration over initial fields from the quantum path integral.
This subtraction is essential to justify the classical approximation,
and it introduces automatically the counterterms for the linear divergences
of the classical theory that otherwise have to be introduced by a rather 
ad hoc matching procedure \cite{iancu}.

The outline of this paper is as follows.
In section \ref{sectwo} we separate off the quantum fluctuations and define 
an influence action containing the quantum corrections to the classical action.
We give the Feynman rules to calculate this action and use it to  
define the thermal weight and effective equations of motion of the effective 
classical theory.
In section \ref{secthree} we calculate the leading order quantum corrections 
to the thermal weight. We consider especially the terms 
\(({\cal E}^{l})^2,{\cal E}^{l}{\cal D}^{l}{\cal A}^{0},
({\cal D}^{l}{\cal A}^{0})^2\) and \(({\cal A}^{0})^2\), since their 
coefficients may be different in the Hamiltonian and
Lagrangian formulations of static dimensional reduction. The leading order 
contribution is the Debye mass term, the only static HTL. We then argue that 
the HTL contributions should be included in the effective equations of motion.
The HTL's obtained through
integrating out the quantum fluctuations are linearly divergent. 
These are, in fact, the counterterms for the linear divergences 
encountered in the effective
classical theory. In section \ref{secfour} we present the effective classical 
theory and discuss the classical approximation.
Section \ref{secfive} contains the conclusion.

\section{Separating off the quantum fluctuations}\label{sectwo}
\subsection{Background fields and zero modes}
In the real-time formulation of thermal field theory, the thermal state 
is taken into account by extending the time evolution along a contour \(C\) 
in the complex \(t\)-plane, as shown in fig. 1. 
The contour consists of three 
parts: 
a forward time branch $C_1$ on the real axis starting at an initial time
$t_{\rm in}$ and ending at a final time \(t_{\rm f}\), 
a backward branch $C_2$ from \(t_{\rm f}\) to \(t_{\rm in}\)
and a third branch $C_3$ down the Euclidean path \(t=t_{\rm in}-i\tau\) 
from \(\tau=0\) to \(\tau=\beta\). On this contour we consider a SU($N$) 
gauge theory for the field 
\(A^{\mu}=(A^{0},-\vec{A})\) 
with generating functional \cite{faddeev,itzykson}, 
\be
Z[j]=\int \D A\D E \D\bar{c}\D
c \;\delta(\vec{\nabla}\cdot\vec{A})
\exp\left({iS[A,\vec{E}]+iS_{\rm gh}[c,\bar{c},\vec{A}]
+ij\cdot A}\right)\;,
\label{genfun}\ee
where \(\vec{E}\) is the momentum conjugate to \(\vec{A}\).
The \(\delta\)-function enforces the Coulomb gauge 
\(\vec{\nabla}\cdot\vec{A}=0\).
We use the 
inner-product notation 
$j \cdot A=-2 {\rm tr}
\int_{C}d^4x\,j_{\mu}(x)A^{\mu}(x)$; the 
notational conventions are those of reference \cite{itzykson}. 
The action of the theory is
\be
S[A,\vec{E}]= 2 {\rm tr}\int_{C}d^4 x\, \left[\vec{E}\cdot\partial_{t}\vec{A}+
\half\left(\vec{E}^{2}+ \vec{B}^{2} \right)-A^{0}R\right]\;.
\label{action}
\ee
Here \(\vec{B}\) is the magnetic field and \(A^{0}\) 
is the Lagrange multiplier associated with the constraint
\(R(A,E):=\vec{\nabla}\cdot\vec{E}+g[\vec{A},\vec{E}]=0\); it is real 
on the real-time contour and complex on the imaginary time-branch
\(A^{0}(t_{\rm in}-i\tau,\vec{x})= iA^{4}(\tau,\vec{x})\).  
The time-derivative of the field \(\vec{A}\) in (\ref{action}) 
is taken along the contour. The integration over the ghost action 
\be
S_{\rm gh}= -2\mbox{tr}\int_{C}d^{4}x \, \left(
\bar{c}\nabla^2 c-g \bar{c}\;[\vec{A},\vec{\nabla}c]\right)
\label{ghostaction}
\ee 
generates the Faddeev-Popov determinant. The fields \(\vec{A}\) 
and the ghosts are 
periodic with respect to begin- and endpoint of the contour, but the momenta 
\(\vec{E}\) and the Lagrange multiplier \(A^{0}\) have no boundary conditions. 

We introduce the classical background fields on the real-time contour 
\(C_{12}=C_{1}\cup C_{2}\) by shifting the fields
$A\rightarrow A_{\rm cl}+A, \; 
\vec{E}\rightarrow \vec{E}_{\rm cl}+\vec{E}$, 
and similarly for the ghost fields. Performing
this shift in the action (\ref{action}), we obtain
\be
S[A_{\rm cl}+A,\vec{E}_{\rm cl}+\vec{E}]=S[A,\vec{E}]_{\rm cl}
+\delta_{A}S[A,\vec{E}]_{\rm cl}\cdot A
+S[A_{\rm cl},\vec{E}_{\rm cl};A,\vec{E}]\;,
\label{shiftedaction}
\ee
where we have defined 
$S[A,E]_{\rm cl}= \left.
S[A,E]\right|_{A=A_{\rm cl},E=E_{\rm cl}}$.
The last term in (\ref{shiftedaction}) is the part of the shifted action 
containing terms quadratic-and-higher in the quantum fluctuations.
The term linear in \(E\)
has  been made to vanish by imposing on \(C_{12}\) the condition
$\delta_{E}S[A,E]_{\rm cl}=0$ which defines the classical electric 
field in term of the potentials
\be
E_{\rm cl}^l
=\partial^{l} A_{\rm cl}^0-\partial^{0} A_{\rm cl}^{l}
-g[A_{\rm cl}^{l},A_{\rm cl}^0]
=: F_{\rm cl}^{l0}\; .
\label{eqmot1}
\ee
on \(C_{12}\). In section \ref{secfour}, we will require the background field 
\(A_{\rm cl}\)
to satisfy the hard thermal loop (HTL) equations of motion.

The procedure for treating the ghost action (\ref{ghostaction}) is the 
same as for the gauge field action (\ref{shiftedaction}). Linear terms 
are eliminated by imposing
$\delta_{\bar{c}}S_{\rm gh}[c,\bar{c},A]_{\rm cl}
=0,\;\delta_{c}S_{\rm gh}[c,\bar{c},A]_{\rm cl}=0$.
Since the ghost fields are not dynamical in the Coulomb gauge, this implies
\(c_{\rm cl}=0,\;\bar{c}_{\rm cl}=0\).

On the Euclidean part of the contour we follow the prescription of static 
dimensional reduction and split the fields into zero and non-static 
modes. The zero modes are constant in time and defined by a projection of 
the fields on \(C_{3}\); for example
\be
{\cal A}(\vec{x})={\cal P}A(t,\vec{x})=i T\int_{C_{3}}dt A(t,\vec{x})\;.
\label{projection}\ee
The other zero modes are defined similarly and denoted as \({\cal E},
{\cal C},\bar{{\cal C}}\). The static and the non-static modes are separated 
in the integration measure by formally writing
$DA\rightarrow D{\cal A}\left[DA\delta\left({\cal P}A-{\cal A}\right)\right]
\;$,
and similarly for the other fields. Since in the integration over the quantum 
fields \(A\) the zero mode \({\cal A}\) is treated as a constant background 
field, we may shift \(A\rightarrow A+{\cal A}\) like on the contour \(C_{12}\).
The shifted action has the same form as in (\ref{shiftedaction}) if we identify
\(A_{\rm cl}(x)={\cal A}(\vec{x})\), 
\(E_{\rm cl}(x)={\cal E}(\vec{x})\),
\(C_{\rm cl}(x)={\cal C}(\vec{x})\) and 
\(\bar{C}_{\rm cl}(x)=\bar{{\cal C}}(\vec{x})\) on $C_{3}$, 
except that there is no 
linear term. This is a well known feature of static dimensional reduction.
In the present case one may verify that the linear terms vanish on \(C_{3}\), 
because they contain the projections \({\cal P}A\), etc.

\subsection{Influence action}
Performing the  manipulations described in the previous section, 
we separate the generating functional (\ref{genfun}) into a path
integral over static fields
\bea
Z[j]&=&\int \D{\cal A}\D{\cal E}\D{\cal C}\D\bar{{\cal C}}
\delta (\vec{\nabla}\cdot\vec{\cal A})\nn\\
& &\exp\left(iS[A_{\rm cl},\vec{E}_{\rm cl}]
+ iS_{\rm gh}[c_{\rm cl},\bar{c}_{\rm cl},\vec{A}_{\rm cl}]
+ iW[A_{\rm cl},\vec{E}_{\rm cl}, c_{\rm cl}, \bar{c}_{\rm cl}; J]+
ij \cdot A_{\rm cl}\right)\;,
\label{genfun2}
\eea
and one over the quantum fields which defines the influence action
\bea
\exp{iW[A_{\rm cl},\vec{E}_{\rm cl},c_{\rm cl},\bar{c}_{\rm cl};J]}
&=&\int \D A_{\mu}\D c
\D\bar{c}
\delta\left({\cal P}A\right)
\delta\left({\cal P}c\right)\delta\left({\cal P}\bar{c}\right)
\delta (\vec{\nabla}\cdot\vec{A})\nn\\
& &\exp\left({iS[A_{\rm cl},\vec{E}_{\rm cl};A]
+ iS_{\rm gh}[c_{\rm cl},\bar{c}_{\rm cl},A_{\rm cl};c,\bar{c},A]
+iJ \cdot A}\right)\;.
\label{effact}
\eea
Here we used that the gauge constraint may be imposed separately on the zero 
and non-static modes. 
The source \(J\) is the combination of the external source and the linear term
in (\ref{shiftedaction}) on \(C_{12}\)
\be
J_\mu =j_\mu +\delta_{A^\mu}S_{\rm cl} \; .
\label{source}\ee
The ghost action 
\(S_{\rm gh}[c_{\rm cl},\bar{c}_{\rm cl},\vec{A}_{\rm cl};c,\bar{c},\vec{A}]\) 
contains the terms quadratic-and-higher in the quantum fluctuations
arising from the shifted ghost action. The Gaussian integration over the 
momenta 
$\vec{E}$ has already been performed, 
which reduces the action in the exponent of (\ref{effact}) to the expression
\be
S[A_{\rm cl},\vec{E}_{\rm cl};A]= -2 \,\mbox{tr}\int_{C}d^4 x\, \left(
-\frac{1}{4}F_{\mu\nu}F^{\mu\nu}+
\half g[A^{k},A^{l}]\,F^{kl}_{\rm cl}-gA^{0}\,[A^{l},E^{l}_{\rm cl}]
-Q \right)\;,
\label{fluctaction}
\ee 
with the classical field strength
\(F_{\rm cl}^{kl}=
\partial^{k}A^{l}_{\rm cl}-\partial^{l}A^{k}_{\rm
cl}-g[A^{k}_{\rm cl},A^{l}_{\rm cl}]\), \(k,l=1,2,3\).
The quantum field strength tensor 
\be
F^{\mu\nu}=\partial^{\mu}A^{\nu}-\partial^{\nu}A^{\mu}-g[A^{\mu}_{cl},A^{\nu}]-g[A^{\mu},A^{\nu}_{cl}]-g[A^{\mu},A^{\nu}]\;,
\label{fieldstrength}\ee
contains linear contributions from the background field.  
The last term in (\ref{fluctaction}) is given by
\be
Q=\half({\cal P}F^{0l})^2\;,
\label{Q}\ee
arises from the integration of the
momenta and lives only on \(C_{3}\).

We note that on the real-time contour \(C_{12}\) the action 
(\ref{fluctaction}) is just the standard quantum action for SU($N$)  
shifted by a classical background: 
\be
S[A_{\rm cl},\vec{E}_{\rm cl};A]+J\cdot A=S[A_{\rm cl}+A]-S[A_{\rm cl}]+j\cdot
A\; . 
\label{checkC12}\ee
To derive (\ref{checkC12}) we inserted the equations of motion 
(\ref{eqmot1}) for 
$\vec{E}_{\rm cl}$  into the action (\ref{action}), which gives the Lagrangian 
version  
$S[A]:=S[A,E^{l}=F^{l0}]$ for  the SU($N$) action. 
On \(C_{3}\) the electric field 
$\vec{E}_{\rm cl}=\vec{\cal E}$ is an independent integration variable and 
we cannot
simplify to a form like (\ref{checkC12}).

\subsection{Feynman rules}\label{secFeyn}
The effective action can be calculated perturbatively with the Feynman rules 
determined by the action (\ref{fluctaction}). 
By inspection of (\ref{checkC12}) we see that on \(C_{12}\) the interaction 
vertices are the usual vertices with two, three or four quantum fields on the 
external lines. Hence, the Feynman rules for 
a diagrammatic evaluation  of vertex functions on 
\(C_{12}\)  consist of the usual interactions.
But the propagator of the quantum fluctuations is not the usual propagator 
of thermal field theory, as we will derive below.

On \(C_{3}\) there is an interaction containing the static 
electric field, this  interaction gives corrections, for instance, to the 
\(\vec{\cal E}^2\) term in the classical action. There are also interactions 
contained in \(Q\) (\ref{Q}) these do not contribute at one-loop order and are 
not considered in this paper. However the quadratic part of \(Q\) is however 
essential in the computation of the propagator of the quantum fluctuations.

The influence action (\ref{effact}) for free fields only, depends on 
the background fields through the source (\ref{source}) and will be denoted 
by $W_{0}[J]$. This quantity determines the propagator $\Delta^{\mu\nu}$ 
associated with the quantum fluctuations according to 
\be
\exp iW_{0}[J]=\exp-\half \int_{C}dtdt'\int
\frac{d^{3}k}{(2\pi)^3} \tilde{J}_{\mu}(t,-\k)
\tilde{\Delta}_{C}^{\mu\nu}(t-t',\k) \tilde{J}_{\nu}(t',\k)\; ,
\label{defprop}
\ee
where the tilde indicates the spatial Fourier transform. 
To evaluate the left-hand-side we insert an integral representation
\(\delta\left({\cal P}A\right)=
\int \D \chi_{\mu}\exp\left(-\chi\cdot A\right)\)
for the delta-function in (\ref{effact}). The auxiliary field $\chi$ is zero 
everywhere except on \(C_{3}\) where 
it has a spatial dependence: \(\chi_{\mu}(t,\vec{x})=\chi_{\mu}(\vec{x})\;  , 
t\in C_{3}\). 
The quadratic part of the term \(Q\), equation (\ref{Q}), reduces to
$Q_{0}=({\cal P}\partial_{t}\vec{A})^2\;$,
since \({\cal P}A^{0}\) vanishes on account of the delta-function constraints 
in (\ref{effact}). These manipulations allow us to rewrite the free effective 
action in the form
\be
\exp iW_{0}[J]=\exp\left(\frac{i}{2}{\cal P}\partial_{t}\delta_{J}\cdot
{\cal P}\partial_{t}\delta_{J}\right)
\int \D_{C} \chi_{\mu}\exp
-\half (J+i\chi)\cdot
D_{C} \cdot(J+i\chi)\; ,
\label{quadreffact}\ee
with \(D_{C}^{\mu\nu}\) the standard thermal propagator of gauge fields in the 
Coulomb gauge \cite{lebellac}. The contribution of the \(Q\)
term has been taken out of the functional integration by replacing the 
spatial gauge fields 
by the appropriate functional derivatives. 

The propagator $\tilde{\Delta}_{C}^{\mu\nu}$ has the same tensor structure 
as the standard thermal propagator in the Coulomb gauge:
\be
\tilde{\Delta}_{C}^{\mu\nu}(t-t',\vec{k})=
T^{\mu\nu}(\vec{k})\tilde{\Delta}_{C}(t-t',\vec{k})
+g^{\mu 0}g^{\nu 0}\tilde{\Delta}_{C}^{00}(t-t',\vec{k})\;.
\label{qprop}
\ee
The matrix \(T\) is the transversal projection operator 
with \(T_{0\nu}=T_{\mu 0}=0\) and
\be
T_{ij}=\delta_{ij}-\frac{k_{i}k_{j}}{k^2}\;,
\ee
with \(k=|\vec{k}|\).
The evaluation of the temporal part of the propagator (\ref{qprop}) is simple,
since the functional derivatives in (\ref{quadreffact}) do not contribute 
and the  thermal propagator \(\tilde{D}_{C}^{00}\) is local in time:
\be
\tilde{\Delta}_{C}^{00}(t-t',\vec{k})=\tilde{D}_{C}^{00}(t-t',\vec{k})-
\tilde{S}_{C}^{00}(t-t',\vec{k})\;,
\label{prop00}\ee
with subtraction
\be
\tilde{S}_{C}^{00}(t-t',\vec{k})=\left\{
\begin{array}{ll}
0& \quad t\mbox{ or }t'\in C_{12}\;,\\
i \frac{T}{k^2}& \quad t,t'\in\; C_{3}\; .
\end{array}\right.
\label{subtr00}
\ee
The calculation of the scalar propagator \(\tilde{\Delta}_{C}\) for the 
spatial gauge fields is more involved; we refer to \cite{nauta} for details.
The result is
\be
\tilde{\Delta}_{C}(t-t',\vec{k})=
\tilde{D}_{C}(t-t',\vec{k})-\tilde{S}_{C}(t-t',\vec{k}),
\label{prop}
\ee
with \(\tilde{D}_{C}\) the standard thermal propagator \cite{lebellac}. The 
subtraction is given by \cite{nauta}
\be
\tilde{S}_C(t-t',\vec{k})=\left\{
\begin{array}{ll}
\frac{T}{k^2}\cos k(t-t')& \quad t,t'\in C_{12},\\
\frac{T}{k^2}\cos k(t_{\rm in}-t')& \quad t\in C_{3}, t'\in C_{12},\\
\frac{T}{k^2}\cos k(t- t_{\rm in})&\quad t\in C_{12}, t'\in C_{3},\\
\frac{T}{k^2}& \quad t,t'\in C_{3}.
\end{array}
\right.
\label{clasprop}
\ee
On the real-time part of the contour we recognize the subtraction as the 
classical propagator. On the vertical part of the contour it is the zero-mode 
propagator, which
implies that static fluctuations on \(C_{3}\) do not contribute 
to the propagator (\ref{qprop}). This is what one expects, 
of course, since static fluctuations are excluded from the path 
integration in (\ref{effact}). 

The propagator for the ghosts can be derived in the same way as for the gauge 
fields. In the Coulomb gauge it  is equal to the propagator for the temporal 
fluctuations
\be
\tilde{\Delta}_{C}^{\rm gh}(t-t',\vec{k})=\tilde{\Delta}_{C}^{00}(t-t',\vec{k})\; . 
\label{propgh}\ee

The important conclusion is
that the IR-limit of the propagator $\tilde{\Delta}_{C}^{ij}$ for the 
transverse quantum fluctuations is ${\cal O} (k^{-1})$ and, therefore less 
singular than the IR-limit ${\cal O} (k^{-2})$ of the standard thermal 
propagator $\tilde{D}_{C}^{ij}$. 
So the severe IR divergences that are present in 
massless thermal field theories in perturbation theory do not occur in a 
perturbative evaluation of the influence action.
What we have achieved is a clean separation between quantum fluctuations that 
may be treated perturbatively, and non-perturbative effects contained in the 
3D path integral (\ref{genfun2}).  

\subsection{Thermal weight and equation of motion}\label{secthermeq}

The classical fields are defined on the entire contour \(C\).
The classical action \(S\) and the effective action \(W\) in
(\ref{genfun2}), contain contributions from classical fields on the
entire contour. The contributions from \(C_3\) and \(C_{12}\)
differ completely: the former contributes to the thermal weight
whereas the latter determines the time evolution. This difference
may be made more explicit by demanding
\be
\vec{A}_{\rm cl}(t,\vec{x})=\vec{A}_{\rm cl}(t-i\sigma,\vec{x})\;,
A^{0}_{\rm cl}(t,\vec{x})=-A^{0}_{\rm cl}(t-i\sigma,\vec{x})\;,
\label{demand}\ee
with \(t\in C_{1}\) and \(t-i\sigma\in C_{2}\).
Then (\ref{eqmot1}) implies
\(
\vec{E}_{\rm cl}(t,\vec{x})=-\vec{E}_{\rm cl}(t-i\sigma,\vec{x})
\). 
We recall that the temporal gauge field and the electric field need not be 
periodic on the entire contour, so the minus sign in (\ref{demand}) 
does not cause problems.
The consequence is that
the contributions to the classical action from the forward and
backward time-branch cancel and the contribution
from the imaginary time-branch \(C_3\) remains
\be
iS_{\rm cl}=-\beta H[\vec{\cal A},\vec{\cal E}]+\beta {\cal A}^{0}{\cal R}\;,
\label{redclasact}\ee
with the abbreviation
\({\cal A}^{0}{\cal R}=
-2\mbox{tr}\int d^3x{\cal A}^{0}R[\vec{\cal A},\vec{\cal E}]\),
and 
\be
iS_{\rm gh}[c_{\rm cl},\bar{c}_{\rm cl},\vec{A}_{\rm cl}]=
iS_{\rm gh}[{\cal C},\bar{{\cal C}},\vec{\cal A}]\;,
\label{redghostact}\ee
for the ghost action.
The effective action may be split into a part containing only
contributions from fields on \(C_3\) and a part containing
contributions from fields both on \(C_{12}\) and \(C_3\)
\be
W[A_{\rm cl},\vec{E}_{\rm cl},c_{\rm cl},\bar{c}_{\rm cl},J]=
W_{\rm DR}[{\cal A},\vec{\cal E},{\cal C},\bar{{\cal C}}]+
W_{\rm RT}[A_{\rm cl},\vec{E}_{\rm cl},c_{\rm cl},\bar{c}_{\rm cl},J]
\label{redeffact}\ee
with \(W_{\rm DR}\) the influence action of static dimensional reduction
and the real-time influence action \(W_{\rm RT}\).
The different roles plaid by these two actions 
becomes clear by noting that \(W_{\rm DR}\) is source independent, while 
a consequence of (\ref{demand})
is that the real-time action vanishes at zero source 
\(\left.W_{\rm RT}\right|_{j=0}=0\), as for scalar theories \cite{nauta1}.
Hence, the influence action of static dimensional reduction gives the quantum 
corrections to the thermal weight, while the real-time
influence action modifies 
the source term.

Using the identities (\ref{redclasact}), (\ref{redghostact}) and
(\ref{redeffact}) for the generating functional (\ref{genfun2}), we may write
\bea
Z[j]&=&\int\D {\cal A}\D{\cal E}\D{\cal C}\D\bar{{\cal C}}
\delta(\vec{\nabla}\cdot\vec{\cal A})
\exp\left(-\beta H[\vec{\cal A},\vec{\cal E}]+\beta {\cal A}^{0}{\cal R}
+iS_{\rm gh}[{\cal C},\bar{\cal C},\vec{\cal A}]+iW_{\rm DR}\right)\nn\\
& &\;\;\;\;\;\;\;\;\;\;\;\;\;\;\;\;\;\;\;\;\;\;
\exp\left( iW_{\rm RT}+ij\cdot A_{\rm cl}\right)\;,
\label{redgenfun}\eea
where the first exponent is the effective thermal weight and the second 
exponent is the effective source term.

We still have to make a choice for \(A_{\rm cl}\)
on the real-time part of the contour. The generating functional
is independent of this choice and the idea is to minimize
the quantum corrections. 
This may be implemented by demanding
\be
\delta_{j}\left.
W_{\rm RT}[A_{\rm cl},\vec{E}_{\rm cl},c_{\rm cl},\bar{c}_{\rm cl};J]\right|_{j=0}=0\;.
\label{demandcl}\ee
To rewrite this requirement as an equation of motion for the classical field, 
we define the effective action
\be
\Gamma[A_{\rm cl},\tilde{A}]= 
W_{\rm RT}[A_{\rm cl},\vec{E}_{\rm cl},c_{\rm cl},\bar{c}_{\rm cl};J]
-J\cdot\tilde{A}\;,\;\;\;\delta_{J}W_{\rm RT}=\tilde{A}
\label{Legtran}\ee
where \(\tilde{A}\) is a general fields on \(C_{12}\) (not necessarily 
equal on \(C_1\) and \(C_2\)). The effective action depends on the zero modes 
\({\cal A},\vec{\cal E},{\cal C},\bar{\cal C}\) (not explicitly shown)
and the field \(\,\tilde{A}\) on \(C_{12}\).
Given a source \(J\) the field \(\tilde{A}\) satisfies
\be
\left.\delta_{A}\Gamma[A_{\rm cl},A]\right|_{A=\tilde{A}}=-J\;.
\label{eqtila}\ee
Imposing (\ref{Legtran}) this gives an equation for the classical field on 
\(C_{12}\)
\be
\left.\delta_{A}\Gamma[A_{\rm cl},A]\right|_{A=0}=
-\left.\delta_{A}S_{12}\right|_{\rm cl}\;.
\label{effeqmota}\ee
We introduce \(\Gamma[A]:=\Gamma[A,\tilde{A}=0]\), where we allow for a
general field \(A\) on \(C_{12}\).
Using the identity \(\left.\delta_{A}\Gamma[A_{\rm cl},A]\right|_{A=0}=
\left.\delta_{A}\Gamma[A]\right|_{{\rm cl}}\) (which follows immediately 
from a generalization of (6.4) in \cite{fujimoto}), we may write  (\ref{effeqmota}) as
\be
\left.\delta_{A}
\left(S_{12}[A]+\Gamma[A]\right)
\right|_{{\rm cl}}=0\;.
\label{effeqmot}
\ee
The action \(\Gamma\) contains the loop corrections to the classical equations
of motion. These loop corrections are calculated with the propagator \(\Delta\)
on the internal lines.

Note that we may set \(j=0\) in (\ref{demandcl}) and (\ref{effeqmot}),
since we have initial conditions on the classical fields at a finite time 
\(t_{\rm in}\) and therefore (\ref{effeqmot}) is not a trivial requirement.
This in contrast with  situations where one has the asymptotic condition
\(A_{\rm cl}\rightarrow 0\), and it is necessary to keep the source in order
to obtain non-trivial solutions.
 
A useful property of the vertex functions defined by \(\Gamma[A]\)
is that they satisfy the 
 Slavnov-Taylor identities on \(C_{12}\). 
This can be seen by noting that setting the 
classical fields equal to zero on \(C_{12}\) gives the standard action 
for the gauge fields (\ref{checkC12}) and the ghost fields. Therefore the 
action is invariant under BRS transformations of the quantum fields 
\(A\). Also the integration measure \(\D A\delta\left({\cal P}A\right)
\D c\delta\left({\cal P}c\right)\D \bar{c}\delta\left({\cal P}\bar{c}\right)\)
is invariant under a BRS transformation on \(C_{12}\), since \({\cal P}\)
does not act on \(C_{12}\). From the BRS invariance 
the Slavnov-Taylor identities follow in a standard manner.

\section{Quantum corrections  }\label{secthree}
\subsection{Thermal weight}
In this section we use the previously derived Feynman rules to
calculate the leading-order quantum corrections to the 
thermal weight in the generating functional (\ref{redgenfun}).

Let us assume we may write the effective action as an infinite sum of 
local terms, symbolically
\be
iW_{\rm DR}=2\beta \;\mbox{tr}\int d^3 x \sum K (\partial)^{P}({\cal A})^{Q}
({\cal E})^{R}({\cal C})^{S}(\bar{{\cal C}})^{S}\;,
\ee
where the ghost and anti-ghost fields have the same power since these are
Grassmann fields.
The factor \(\beta\) arises from the imaginary time integration over \(C_3\).
The above equation is a derivative expansion and is valid if the momenta of 
the fields are small \(k<<T\). By power counting and a one-loop estimate
we
find for the coefficient 
\be
K\sim g^{Q+R+2S} T^{4-N}\;,
\label{coefficient}\ee
with \(N=P-Q-2R-2S\). 
We see that terms with \(4-N<0\) are suppressed 
for small momenta \(k<<T\) and small fields 
\(g{\cal A},g{\cal C}<<T,\; g{\cal E}<<T^2\) \cite{kajantie}.
Note also that the superficial divergence structure is given by the same 
power counting: \(K\) is superficially finite if \(4-N<0\).
 
We are interested in the leading quantum corrections 
and we concentrate on terms with \(4-N\geq 0\).
Since the theory is invariant under static BRS transformations
the relevant terms in the effective action must have the form
\bea
iW_{\rm DR}[{\cal A},\vec{\cal E},{\cal C},\bar{{\cal C}}]=2\beta\mbox{tr}\int d^3x
& &\left[\frac{1}{4}K_1{\cal F}^{kl}{\cal F}^{kl}
+\half K_2 {\cal E}^{l}{\cal E}^{l}
-K_3 {\cal E}^{l}{\cal D}^{l}{\cal A}^{0}
+\half K_{4}\left({\cal D}^{l}{\cal A}^{0}\right)
\left({\cal D}^{l}{\cal A}^{0}\right)\right.\\\nn
& &\left.-\half m_{0}^2 ({\cal A}^{0})^2
+\frac{1}{4!}\lambda_{0}({\cal A}^{0})^4
+K_{5}\left(\bar{{\cal C}}\partial^{k}\partial^{k}{\cal C}
-g \bar{{\cal C}}[A^{k},\partial^{k}{\cal C}]\right)+...\right]\;,
\label{effactC3}\eea
with \({\cal F}^{kl}\) and \({\cal D}^l\) the static field strength
and covariant derivative, respectively. 
This is the Hamiltonian form of the effective action as found in
\cite{landsman,kajantie}.

The coefficients \(K_1...K_5,m_0^2\) and \(\lambda_0\) in (\ref{effactC3}) 
may be calculated 
in perturbation theory, since the subtractions in (\ref{prop00}),(\ref{prop}) and 
(\ref{propgh}) imply that the perturbative expansion is not plagued with 
the severe IR divergences of thermal field theory.

The mass for the temporal gauge field equals the well known result for the
Debye mass
\be
m_{0}^2=-2g^2 N \mu^{2\epsilon}\int\frac{d^{d}k}{(2\pi)^d}\frac{d}{dk}
\tilde{n}(k)=\frac{1}{3}g^2 N T^2\;,
\label{debyemass}\ee
with \(k=|\vec{k}|\). We use dimensional regularization 
to render the divergent (spatial) momentum integrals finite;
 \(d=3-2\epsilon\), \(\mu\) is a reference mass.
The distribution function of the
quantum fluctuations is 
\(
\tilde{n}(k)=n(k)-T/k
\), and
comes from the thermal part of the propagator (\ref{prop}).
The subtraction \(T/k\) introduces a linear divergence in
the mass (\ref{debyemass}). This divergence is temperature dependent
in contrast to the 4D divergences of quantum field theory, 
and acts as a counterterm for the 
same divergence arising in the dimensionally reduced theory.
In dimensional regularization power divergences are set to zero
and the subtraction does not contribute to the mass. Hence the result is the
leading order contribution to the Debye mass.

The other constants appearing in (\ref{effactC3}) are dimensionless 
\(4-N=0\). We find
\bea
K_2&=&K_3=-\frac{1}{3} g^2 N\mu^{2\epsilon}
\int \frac{d^dk}{(2\pi)^d}
\frac{1}{k^3}(1+2\tilde{n}(k))-\mbox{c.t.}
=-g^2 N \frac{1}{24\pi^2}\left(\frac{1}{\hat{\epsilon}}+2\right)-\mbox{c.t.}
\;,\nn\\
K_4&=&K_2+g^2N\frac{5}{48\pi^2}\left(\frac{1}{\hat{\epsilon}}+\frac{26}{15}
\right)-\mbox{c.t.}\;,
\label{calcK}\eea
with \(\frac{1}{\hat{\epsilon}}=\frac{1}{\epsilon}+\gamma-
\log\frac{4\pi T^2}{\mu^2}\) and \(\gamma\) Eulers constant
and c.t. the appropriate counterterms. Note that the subtraction in 
\(\tilde{n}\) avoids a linear infrared divergence in \(K_2,K_3\) and \(K_4\).

The constants \(K_1, K_5\) and \(\lambda_0\) have been calculated before in 
the
Lagrangian formulation of dimensional reduction and may be found in
\cite{kajantie,landsman} for covariant gauges. 
Integrating out the static momenta \({\cal E}\)
yields the dimensionally reduced action in the Lagrangian formulation.

Unlike for the linear divergence in (\ref{debyemass}),
temperature-independent 4D counterterms are sufficient to render the
constants \(K_1,...,K_5,\lambda_0\) finite.
The constants \(K_1,...,K_5\) are subleading in \(g\) 
compared to the classical action. The term
\(\lambda_0 (A^0)^4\) is small compared to the term \(m_{0}^2 (A^0)^2\) 
for not too large fields \(A^0\sim T\), since \(\lambda_{0}\sim g^4\); see 
(\ref{coefficient}). Hence, we conclude that only the \(m_{0}^2 (A^0)^2\)  
is not perturbatively small compared to the classical action.
This is the only static HTL and
has to be included in the thermal weight for the effective 
classical theory, while the other subleading contributions can be treated 
as perturbations.

\subsection{Hard thermal loop contributions to the equation of motion}
\label{secHTL}

We now turn to the equation of motion (\ref{effeqmot}). The effective 
action \(\Gamma\) 
contains all quantum effects and is a complicated expression.
However, since we are interested in low-momentum correlation functions
\(k<<T\), we may consider only the leading-order terms to the equation 
of motion at low momenta and weak-coupling. 

The leading contributions to \(\Gamma\) at low momenta are the well-known 
HTL's. We review the order estimates \cite{arnold,braatenpisarski}
for the HTL contributions to the action \(\Gamma\)
and discuss the modifications due to 
the subtraction in the quantum propagator. 

In the effective classical theory one has
two length scales 
\((gT)^{-1}\) and \((g^2T)^{-1}\).
These two scales correspond to two different excitations; the plasmons that
are screened over a length scale \((gT)^{-1}\) \cite{plasmon}, 
and the attenuating modes \cite{lebedev}
that are screened non-perturbatively over a length scale \((g^2T)^{-1}\)
due to a magnetic screening mass in 3D gauge theories \cite{gapeq}.

Let us first consider the momentum and energy scale \(k\sim gT, k_0\sim gT\)
of the plasmon modes.
We assume that the classical fields are of the same order as 
the initial fields \(A_{\rm cl}\sim {\cal A}\). An estimate for the initial 
fields is found by requiring that the fields are not suppressed by
the thermal weight:
\(\beta H\sim 1
\rightarrow {\cal A}\sim \sqrt{g}T\). This yields the 
following estimates for terms in the HTL action
\be
\Gamma_{HTL}^{(n)} A_{\rm cl}^n\sim g^{2+\half n} T^4 \;,
\ee
where \(\Gamma_{HTL}^{n}\) are
the retarded vertex functions, as they appear in the 
equations of motion.
By definition of the HTL's, the classical and HTL contributions are of 
the same order. The interaction terms are small
compared to the quadratic parts and therefore (resummed)
perturbation theory is valid for the plasmon scale \cite{braaten3,schulz}.

Secondly we consider the momentum scale \(k\sim g^2 T\) and energy scale 
\(k_0\sim g^4T\) of the attenuating modes \cite{arnold}. 
These  are only present in the
spatial gauge fields, since the 
temporal gauge field is electrically screened with Debye screening length
\((gT)^{-1}\). Again we assume that the classical fields and initial fields 
are of the same order, yielding 
\(A_{\rm cl}\sim{\cal A}\sim gT\). 
We have the following estimates for 
terms in the classical and HTL action
\be
\Gamma_{HTL}^{(n)} A_{\rm cl}^n\sim g^{2-n} T^{3-n}k_{0} A_{\rm cl}^n\sim g^6T^4\;.
\label{estHTLact}\ee
Again 
the HTL contributions are as large as the classical contributions. We see that 
higher order terms are not small compared to the quadratic part of the action,
which means perturbation theory fails. It will not be possible to expand
in the non-linearities of the effective classical theory. The estimate for 
the classical and HTL action (\ref{estHTLact}) is of the order where the 
perturbation expansion breaks down for the free energy \cite{linde}.

In thermal field theories large contributions to vertex functions may also 
come from the IR-region \cite{linde}. 
These may be as large as the HTL and 
classical contributions. However these contributions are not present in 
(\ref{effeqmot}) since we use
the subtracted propagator, which implies that IR contributions are suppressed 
by a factor \(\beta k\sim g^2\). 
The large IR contributions are included in the effective classical theory.
This leads to the conclusion that the HTL's have to be included in the 
effective equations of motion and that at the scale \(g^2T\) a perturbation 
expansion for the effective classical theory fails \cite{bodeker,iancu}.

Let us consider HTL's of the quantum fluctuations.
These HTL's are calculated with the Feynman rules given
in section \ref{secFeyn}. We focus on the retarded vertex functions, 
since these 
appear in the equations of motion. We may restrict ourselves to vertex 
functions with all (external) times on \(C_{12}\), since also correlations
between the initial fields and the classical fields at later times can be 
expressed in these vertex functions as explained in \cite{nauta1}. Since the 
HTL's consist of one loop diagrams we only need the interactions and quantum
propagator on \(C_{12}\). The interactions are the usual 3- and 4-point 
interactions, as we see from (\ref{checkC12}). The only difference 
with the usual
Feynman rules of thermal field theory
 is in the propagator, where the Bose-Einstein 
distribution function is replaced by a subtracted distribution function
in the quantum propagator
\(n_{BE}(k)\rightarrow\tilde{n}(k)=n_{BE}(k)-T/k\).
However, we shall now argue that this does not affect the well-know HTL 
expressions \cite{braatenpisarski} in dimensional regularization.
Retarded one-loop diagrams contain one
distribution function \cite{braatenpisarski}. 
The momentum integrals of the usual HTL's 
have a cut-off at \(k\sim T\) due to the Bose-Einstein distribution function
yielding a \(T^2\) behavior. The subtraction in the distribution function of the quantum fluctuations introduces 
a linear divergence in the HTL's. 
Since the classical linear divergences arise only at one-loop and the 
generating functional is finite (including 4D counterterms) the linear 
divergence in the HTL's acts as a  (non-local) counterterm for the classical 
theory.
Since we use dimensional regularization, 
which sets linear divergences equal to 
zero, the HTL's of the quantum fluctuations are just the usual HTL's. 
In appendix A, we analyze the linear divergences using a  momentum 
cut-off on the momenta of the initial fields. It turns out that the linear 
divergences enter in the plasmon frequency only. 
Such a linear cut-off dependence has been proposed before from a matching 
argument \cite{iancu,aarts}.
The analysis here shows how such a counterterm 
arises from integrating out the quantum fluctuations.

The HTL's we thus obtain are gauge invariant, just as the usual HTL's. This is
as a consequence of BRS invariance of \(\Gamma\) 
and the HTL approximations \cite{kobes}.

\section{Effective classical statistical theory}\label{secfour}
\subsection{Exact generating functional}
In section \ref{secthermeq} an effective classical statistical theory 
was defined. Following the literature \cite{arnold,bodeker,iancu}, it 
was argued that the HTL contributions should be included in the effective 
Hamiltonian and the equation of motion. We argued also that the HTL's of 
the quantum fluctuations equal the usual HTL's in dimensional regularization.
With this the exact generating functional can be rewritten as
\bea
Z[j]&=&\int \D{\cal A}\D{\cal E}\D{\cal C}\D\bar{{\cal C}}
\delta\left( \vec{\nabla}\cdot\vec{{\cal A}}\right)
\exp -\beta H_{\rm eff}[{\cal A},\vec{\cal E}]
+iS_{\rm gh}[{\cal C},\bar{{\cal C}},\vec{\cal A}]
+iW_{\rm DR}^{\rm cor}\nn\\
& &\;\;\;\;\;\;\;\;\;\;\;\;\;\;\;\;\;\;\;\;\;\;\;\;\;\;\;\;\;\;\;\;\;\;\;\;
\exp iW_{\rm RT}+ij\cdot A_{\rm cl}\;,
\label{genfun3}\eea
where we have defined the effective Hamiltonian
\be
H_{\rm eff}=H-\beta {\cal A}^{0}{\cal R}+
2\mbox{tr}\int d^3x\half m_{0}^2 ({\cal A}^{0})^2\;,
\label{effham}\ee
and the static non-HTL corrections
\be
W^{\rm cor}_{\rm DR}=W_{\rm DR}+
2i\beta\mbox{tr}\int d^3x\half m_{0}^2 ({\cal A}^{0})^2\;.
\label{nonHTLcor}\ee
In appendix B, it is shown this Hamiltonian 
corresponds to the conserved energy of the classical subsystem.

The HTL equation of motion is
\be
[D^{\nu}_{\rm cl},F^{\rm cl}_{\nu\mu}](x)=3\omega_{\rm pl}^2
\int\frac{d\Omega}{4\pi}
v_{\mu}\int_{-\infty}^{t}dt'U_{\rm cl}(x,x')
v_{\nu}F^{0\nu}_{\rm cl}(x')e^{-\epsilon (t-t')}\;,
\label{HTLeqmot}\ee
with $\omega_{\rm pl}^2=g^2 NT^2/9$ the plasmon frequency. 
The angular integration is over the direction of \(\vec{v}\), 
\(|\vec{v}|=1\).  Furthermore, \(x'=(t',\vec{x}-\vec{v}(t-t'))\)
and the parallel transporter
\(U_{\rm cl}(x,x')=\mbox{P}\exp ( -ig\int_{\gamma} dz_{\mu}
A^{\mu}_{\rm cl}(z) )\) , 
with \(\gamma\) a straight line from 
\(x\) to \(x'\). The damping exponential follows from the 
\(\epsilon\)-prescription of thermal field theory \cite{lebellac}.

The HTL equation of motion has been derived by Blaizot and Iancu 
\cite{blazoit}, using a kinetic approach. They have shown the HTL equation
of motion is a linearized Vlasov equation 
(linear in the deviation from equilibrium),
with the r.h.s. of (\ref{HTLeqmot}) an induced source.

The initial conditions for the equation of motion (\ref{HTLeqmot}) are
\be
A_{\rm cl}^{k}(t_{\rm in},\vec{x})={\cal A}^{k}(\vec{x})\;;\;
F^{0k}_{\rm cl}(t_{\rm in},\vec{x})={\cal E}^k(\vec{x})\;.
\label{incond}\ee
at a finite time \(t_{\rm in}\). The Lagrange multiplier \(A^0\) 
is not a dynamical field, so \({\cal A}^0\) does not act as an 
initial condition. 
The path integration in (\ref{genfun3}) accounts for the thermal average over 
the initial conditions (\ref{incond}).

\subsection{Classical approximation}
Note that the generating functional (\ref{genfun3}) is still exact. The 
necessary approximation to obtain an effective classical theory is to neglect 
\(W_{DR}^{\rm cor}\) and \(W_{RT}\) in (\ref{genfun3}), which yields
\be
Z_{\rm cl}[j]=\int \D{\cal A}\D{\cal E}\D{\cal C}\D\bar{{\cal C}}
\delta\left( \vec{\nabla}\cdot\vec{{\cal A}}\right)
\exp -\beta H_{\rm eff}[{\cal A},\vec{\cal E}]
+iS_{\rm gh}[{\cal C},\bar{{\cal C}},\vec{\cal A}]
+ij\cdot A_{\rm cl}\;.
\label{genfun4}\ee
Let us argue why \(W_{DR}^{\rm cor}\) and \(W_{RT}\) in (\ref{genfun3})
may be neglected for the calculation of correlation functions at low momenta.
The action \(W_{\rm DR}^{\rm cor}\) contains only the non-HTL contributions 
because of (\ref{effham}) and (\ref{nonHTLcor}). Since these  
are small 
compared to the classical and HTL contributions in \(H_{\rm eff}\)
the action \(W_{\rm DR}^{\rm cor}\) may be neglected.

The other approximation \(W_{RT}=0\) can be justified as follows.
We consider as an example the symmetric transverse two-point function
\be
\tilde{D}_{T}(t_1-t_2,\vec{k})=\frac{1}{4}\langle A^{i}(t_1,-\vec{k})T^{ij}(\vec{k})
A^{j}(t_2,\vec{k})+(t_1 \leftrightarrow t_2)\rangle\;,
\ee
which can be obtained from the generating functional (\ref{genfun3})
\be
\tilde{D}_{T}(t_1-t_2,\vec{k})=
-\frac{1}{4}T^{ij}\left[\delta_{j_{i}(t_1-i\sigma)}\delta_{j_{j}(t_2)}+
\delta_{j_{i}(t_1)}\delta_{j_{j}(t_2-i\sigma)}\right] Z[j]\;,
\ee
where the \(\sigma\) indicates that the source derivative is taken on the 
backward contour \(C_2\), see fig. 1.
Since the classical fields satisfy the effective classical equation of motion
the contributions involving \(\delta_{j}W_{RT}\) may be neglected. 
We then find that the two-point 
function can be separated in a classical and a quantum contribution
\be
\tilde{D}_{T}(t_1-t_2,\vec{k})=\tilde{D}^{\rm cl}_{T}(t_1-t_2,\vec{k})
+\tilde{D}^{\rm q}_{T}(t_1-t_2,\vec{k})
\ee
with
\bea
\tilde{D}^{\rm cl}_{T}(t_1-t_2,\vec{k})&=&
\half\langle A_{\rm cl}^{i}(t_1,-\vec{k})T^{ij}(\vec{k})
A_{\rm cl}^{j}(t_2,\vec{k})\rangle_{\rm cl}\;,\\
\tilde{D}^{\rm q}_{T}(t_1-t_2,\vec{k})&=&
-\frac{1}{4}\langle \left[\delta_{j(t_1-i\sigma)}\delta_{j(t_2)}+
\delta_{j(t_1)}\delta_{j(t_2-i\sigma)}\right]  W_{RT}
\rangle_{\rm cl}\;.
\eea
The brackets \(\langle...\rangle_{\rm cl}\) denote the classical average over the initial conditions defined by (\ref{genfun4}).
In temporal Fourier space these contributions take the form
\bea
\tilde{D}^{\rm cl}_{T}(k_0,\vec{k})&=&\rho^{\rm cl}(k_0,\vec{k})n_{\rm cl}(k_0)\;,\\
\tilde{D}^{\rm q}_{T}(k_0,\vec{k})&=&\rho^{\rm q}(k_0,\vec{k})
\half \left[1+2\tilde{n}(k_0)\right]
\;,
\label{fullqprop}\eea
with the classical  spectral density
\be
\rho^{\rm cl}(k_0,\vec{k})=2\mbox{Im}\left[
\frac{1}{-(k_0+i\epsilon)^2+\vec{k}^2+\Pi_{R}^{\rm cl}(k_0,\vec{k})}
\right]\;,
\ee
and \(\Pi_{R}^{\rm cl}(k_0,\vec{k})\) the classical retarded self-energy.
A similar expression can be given for the spectral density \(\rho^{\rm q}\)
with self-energy \(\Pi_R^{\rm q}\).
The appearance of \(\tilde{n}=n_{BE}-T/k_0\) in (\ref{fullqprop}),
instead of the Bose-Einstein distribution function, 
is a consequence of the subtraction in the free propagator (\ref{prop}).
In fact, in the free case (\(\Pi_R^{\rm q}=0\)) the two-point function 
(\ref{fullqprop}) corresponds to the free propagator (\ref{prop}).
Note that in the case that the classical fields satisfy the
effective equation of motion (\ref{effeqmot}) exactly 
\(\delta_j W_{RT}\) vanishes identically and as a consequence
the self-energies \(\Pi_{R}^{\rm cl}, \Pi_{R}^{\rm q}\) are equal to the
full self-energy.
We have required that the classical fields satisfy the HTL equation of motion,
in this case the leading contribution to both the self-energies
\(\Pi_{R}^{\rm cl}, \Pi_{R}^{\rm q}\) is the HTL self-energy.
Indeed, the action \(W_{RT}\) contains the HTL self-energy and all other 
HTL vertices. 
However the quantum contribution \(D^{\rm q}_{T}\) to the symmetric 
transverse two-point function can be neglected at small momenta \(|k_0|<<T\), 
since
\be
\half \left[1+2\tilde{n}(k_0)\right]=\frac{k_0}{12T}+{\cal O}(k_0^2/T^2) << \frac{T}{k_0}\;,\;\;\;\;\;\;\;\;\;\; 0<k_0<<T\;.
\ee
Therefore we may neglect the action \(W_{RT}\) for the calculation of 
(symmetric) correlation functions at low momenta. Note that the subtraction 
in the two-point function for the quantum fluctuations is essential 
to justify the classical approximation (\ref{genfun4}) for the calculation of 
low-momenta correlation functions.

\section{Conclusion}\label{secfive}
We have derived an effective classical theory by integrating out quantum 
fluctuations. The effective equation of motion (\ref{effeqmot}) is
is BRS invariant, which in the HTL approximation implies gauge invariance.
The Feynman rules for the calculation of the effective equation of motion are 
the same as usual except for a subtraction in the propagator. At real-times 
(on the contour \(C_{12}\)) this subtraction enters only in the distribution 
function \(\tilde{n}=n_{BE}-T/k\). We would like to stress that this 
subtraction is a direct consequence of the extraction of the path integration 
over initial fields from the quantum path integral. 
This subtraction is essential in two respects. It allows one to argue that 
the corrections from \(W_{RT}\) to symmetric correlation functions are small 
and therefore the classical approximation is justified.
Secondly, the subtraction introduces linear divergences in the HTL equations 
of motion and the effective Hamiltonian that act as counterterms for the 
linear divergences in the classical theory.
We used dimensional regularization which sets linear divergences equal to zero.
Since a non-perturbative treatment of the classical theory requires a lattice 
implementation, the generalization to a lattice regularization
is important, see \cite{nauta2}.

There are still logarithmic divergences present in the effective classical 
theory (\ref{genfun4}). Therefore the classical theory (\ref{genfun4})
has to be formulated with some cut-off \(\Lambda\), which has to be large 
compared to the plasmon scale \(gT<<\Lambda\). Since logarithmic divergences are suppressed by a factor \(k/T\sim g,g^2\) (at one-loop) or \(g^2\) 
(at two-loops)
the cut-off may be taken as large as the temperature \(\Lambda\sim T\), since 
then the requirement \((k/T)\log(\Lambda/gT), g^2\log(\Lambda/gT)<<1\) is 
fulfilled.

In principle, \(\Gamma\) and \(W_{\rm DR}\) furnish the logarithmic 
counterterms. If these can be extracted explicitly the result is
a cut-off independent classical approximation. 
The counterterms for the one-loop divergences of the classical theory
may be extracted from the one-loop logarithmic divergent contributions in
\(\Gamma\) and \(W_{\rm DR}\). However, it is unclear how one obtains the 
counterterms for the two-loop logarithmic divergences. Yet, we know that
the full one-loop contributions together with the two-loop logarithmic 
divergent contributions should be sufficient.
This suggests that  
an analysis as in \cite{jakovac} for static dimensional 
reduction may be generalized successfully.

\acknowledgments
We would like to thank G. Aarts and J. Smit for useful discussions.

\appendix

\section{Hard thermal loops with a sharp cut-off}
To explicitly see the linear divergence in the HTL's, 
we use a sharp cut-off \(\Lambda\)
on the momenta of the classical 
fields at \(t_{\rm in}\) to regularize divergent momentum integrals.
This imposes the restriction \(k<\Lambda\) on the initial fields and
consequently on the subtraction 
(\ref{clasprop}) on \(C_{12}\).
Repeating the arguments of section \ref{secHTL}, we see that this reduces to a 
replacement
of the Bose-Einstein distribution function of the usual HTL's by
\(\tilde{n}_{\Lambda}\)
\be
n(k)\rightarrow \tilde{n}_{\Lambda}(k)=n(k)-\frac{T}{k}\Theta(\Lambda-k)\;.
\label{cutdistr}\ee
In the HTL expressions for the retarded vertex functions, we use the expansion 
\be
\int dk k^2\,\tilde{n}_{\Lambda}(k+p\cos\Theta)=
\int dk \left[k^2\,\tilde{n}_{\Lambda}(k)-
2k\,\tilde{n}_{\Lambda}(k)p\cos\Theta+...
\right]
\ee
in the external momenta \(p\). The angular integration decouples 
from the \(k\) integration, as usual in the HTL approximation. 
Performing the \(k\) 
integration we find the familiar HTL contributions
to the vertex functions, but with a 
\(\Lambda\)-dependent plasmon 
frequency
\be
(\omega_{\rm pl}^{\Lambda})^2=\omega^2_{\rm pl}-\frac{2}{3\pi^2}g^2 N\Lambda T\;.
\label{omLambda}\ee
This leads to the conclusion that a 
linear divergence appears in the plasmon frequency. 
By the same reasoning as in section \ref{secHTL}, 
it is clear that this linear divergence acts as a counterterm for the 
classical theory.

In the literature \cite{bodeker,BJ,RG}, often a cut-off is introduced 
according to
\be
n(k)\rightarrow n_{\Lambda}(k)=n(k)\Theta(\Lambda -k).
\label{litcut}\ee
In the classical approximation, one chooses \(gT<<\Lambda <<T\)
and approximates the soft 
field by a classical field \cite{bodeker,BJ}.
The classical theory then contains the leading order contributions
to low-momentum correlation functions up to corrections of order
\(\beta \Lambda\).
However, the result including all contributions from integrating out the 
hard modes
is not exact anymore. The reason is that an 
approximation is made on the low-momentum field, namely that it behaves 
classically. 
This implies that the dependence on the cut-off does 
not cancel out in the final result. A linear dependence on the cut-off will 
remain. 

\section{Effective Hamiltonian}

In this appendix we consider the classical subsystem, that is 
the generating (\ref{genfun3})
functional without the quantum corrections
\(W_{\rm DR}^{\rm cor}=0=W_{\rm RT}\).
We show that the effective Hamiltonian in (\ref{effham}) corresponds to the 
energy of the classical subsystem.
We follow Blazoit and Iancu \cite{blazoit} and introduce an
auxiliary field \(W^{0}(x,v)\), that is also a function of the direction
of $\vec{v}$; $v=(1,\vec{v})$ and $|\vec{v}|=1$.
This allows for a local formulation of the equations of motion 
(\ref{HTLeqmot})
\bea
[D^{\nu}_{\rm cl},F^{\rm cl}_{\nu\mu}](x)=3\omega_{\rm pl}^2
\int \frac{d\Omega}{4\pi}v_{\mu}W^{0}(x,v)\label{localeqmot1}\\
\left[v_{\nu}D^{\nu}_{\rm cl},W^{0}(x,v)\right]=F_{\rm cl}^{0\rho}(x)v_{\rho}
\;,\;\;\;\;\;
\label{localeqmot2}\eea
with the covariant derivative 
\(D_{\rm cl}^{\mu}=\partial^{\mu}+igA_{\rm cl}^{\mu}\).
The conserved energy of this system (\ref{localeqmot1},\ref{localeqmot2}) is \cite{blazoit}
\be
E=-\mbox{tr}\int d^3 x \left[\vec{E}^2(x)+\vec{B}^2(x)
+3\omega_{\rm pl}^2\int\frac{d\Omega}{4\pi}W^{0}(x,v)W^{0}(x,v)\right]\;.
\label{energy}\ee
In particular, at \(t_{\rm in}\) the energy is
\be
E=-\mbox{tr}\int d^3 x \left[\vec{\cal E}^2(\vec{x})
+\vec{\cal B}^2(\vec{x})+
3\omega_{\rm pl}^2\int\frac{d\Omega}{4\pi}
W^{0}(t_{\rm in},\vec{x},v)W^{0}(t_{\rm in},\vec{x},v)\right]\;.
\label{inenergy}\ee
The value of the auxilary field at \(t_{\rm in}\) follows from the 
prescription to keep the fields constant in time before \(t_{\rm in}\)
in (\ref{HTLeqmot}). 
In a time-independent background field the vector current is zero
, hence, the vector current at \(t_{\rm in}\) vanishes.
It follows that \(W^{0}(t_{\rm in},\vec{x},v)\)
is independent of the direction of \(\vec{v}\).
A calculation yields
\be
W^{0}(t_{\rm in},\vec{x},v)=A^{0}(t_{\rm in},\vec{x})\;.
\label{w_0}\ee

Using the equation of motion at $t_{\rm in}$, we have
\be
\left[D_{\rm cl}^{l},F^{\rm cl}_{l0}\right](t_{\rm in},\vec{x})=
{\cal R}(\vec{x})=3\omega_{\rm pl}^2 A^{0}(t_{\rm in},\vec{x})\;.
\label{relRA_0}\ee
Inserting (\ref{w_0}) and (\ref{relRA_0}) in (\ref{inenergy}), we find for 
the energy of the classical subsystem
\be
E=-\mbox{tr}\int d^3 x \left[\vec{\cal E}^2(\vec{x})
+\vec{\cal B}^2(\vec{x})+
\frac{1}{3\omega_{\rm pl}^2}{\cal R}^2(\vec{x})\right]\;.
\label{enin}\ee

Let us consider the effective Hamiltonian (\ref{effham}).
Since we neglect the influence actions in (\ref{genfun3}), we can simply
integrate out the zero mode of the temporal gauge field \({\cal A}^{0}\).
This yields the effective Hamiltonian
\(H_{\rm eff}=E\) in the thermal weight.

An effective classical theory based on the equations of 
motion (\ref{localeqmot1},\ref{localeqmot2}), has been proposed by Iancu 
\cite{iancu}.
The average over the initial conditions is taken by a path integration over
\(\vec{\cal A},\vec{\cal E}\) and 
\(W^{0}(t_{\rm in},\vec{x},v)\), constrained by 
Gauss' law.
We see from (\ref{w_0}), that restricting the path integration 
over \(\vec{v}\)-independent auxiliary fields at \(t_{\rm in}\),
is consistent with the inclusion of the correlations between initial 
fields and the fields at later times according to the prescription in 
(\ref{HTLeqmot}). The value for initial auxiliary field is then determined by
Gauss' law (\ref{relRA_0}).
This yields a somewhat simpler effective classical theory than proposed by 
Iancu, since no path integration over \(W^{0}(t_{\rm in},\vec{x},v)\) has 
to be performed.

\newpage
\begin{figure}
\centerline{\psfig{figure=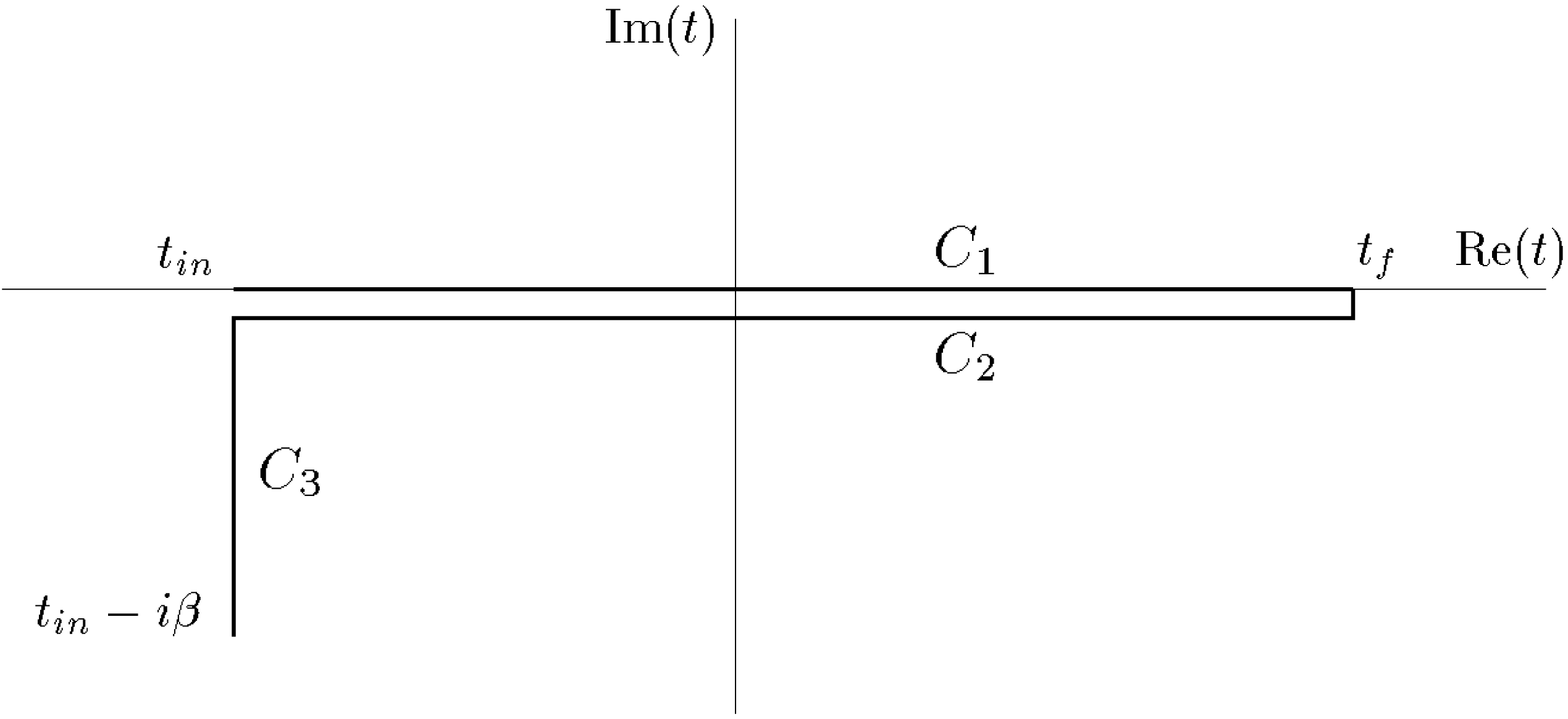,height=2.5in,angle=-90}}
\caption{The time contour \(C\).}
\end{figure}
\end{document}